# Automated Healthcare Thematic Analysis using Multi-Agent Large Language Model: Algorithm Development and Evaluation

Qidi Xu[1], Nuzha Amjad[2], Grace Giles[2], Alexa Cumming[2], De'angelo Hermesky[2], Alexander Wen[2], Min Ji Kwak[2], Yejin Kim[1]*
[1] McWilliams School of Biomedical Informatics, UTHealth Houston, Houston, TX, 77030, USA
[2] McGovern Medical School, UTHealth Houston, Houston, TX, 77030, USA
*Corresponding author:
Yejin Kim
Department of Health Data Science and Artificial Intelligence
McWilliams School of Biomedical Informatics
University of Texas Health Sciences Center Houston
7000 Fannin St. Houston, TX, USA
Yejin.Kim@uth.tmc.edu

## Abstract

Background:
Understanding patients' experiences is essential for advancing patient-centered care, especially in chronic diseases that require ongoing communication. Qualitative thematic analysis is widely used to explore these experiences, however, the process remains labor-intensive, subjective, and difficult to scale.
Objective:
This study aimed to develop and evaluate Collaborative Theme Identification Agent (CoTI), a multi-agent large language model framework designed to support manual thematic analysis by rapidly generating supporting excerpts, initial codes, and themes.
Methods:
CoTI consists of three agents: Instructor, Thematizer, and CodebookGenerator. The Instructor refines instruction prompts, the Thematizer extracts supporting excerpts and generates initial codes for each transcript, and the CodebookGenerator groups similar codes across all transcripts into a codebook with themes. We evaluated CoTI primarily using 12 heart failure patient transcripts focused on perceptions of medication intensity. CoTI-generated outputs were compared against the reference standard developed by senior investigators. To explore human-AI interaction in thematic analysis, we further implemented CoTI in a user-facing application.
Results:
CoTI generated supporting excerpts, initial codes, and themes that were more similar to those of senior investigators than did the outputs of junior investigators, baseline natural language processing models, and other basic large language models. In an exploratory human-AI collaboration experiment, we found that the collaboration between CoTI and junior investigators provided only marginal gains compared to CoTI alone. A possible hypothesis was that junior investigators may over-rely on CoTI and limit their independent critical thinking.

Conclusions:
CoTI can improve the efficiency of thematic analysis by rapidly generating supporting excerpts, initial codes, and themes for human researchers review. These findings highlight CoTI's potential as a useful tool for scalable qualitative research.

## Introduction

Patient-centered care prioritizes understanding and integrating patients' individual needs, values, and preferences into clinical decision-making [1]. This approach is particularly important in the management of chronic diseases such as diabetes, hypertension, and heart failure, which require long-term treatment plans and frequent communication between patients and healthcare providers to ensure timely adjustments in response to changes in the patient's conditions [2,3]. To explore how patients perceive and navigate their health experiences, qualitative research, particularly through thematic analysis of interview transcripts, has been widely employed, offering rich sociocontextual understanding of complex health phenomena [4,5]. For example, previous qualitative studies in heart failure have used thematic analysis to identify themes related to medication intensity and self-management capacity [6,7].

Although thematic analysis has been widely used to generate valuable insights in patient-centered care, it also faces several practical challenges. As a traditional qualitative approach, it typically relies on trained experts to manually review transcripts, extract supporting excerpts (referred to as *clues*), interpret their meanings to identify initial codes, and summarize these codes into a codebook with themes for interpretation (Fig 1a). This process is time-consuming, labor-intensive, and susceptible to subjective interpretation, as experts may differ in how they extract and summarize information. These limitations may introduce variability and potential bias into the findings [8]. To address the challenges of manual thematic analysis, researchers have explored the use of natural language processing (NLP) to assist with thematic analysis [8]. Traditional unsupervised NLP techniques, such as Latent Dirichlet Allocation (LDA) [9], identify recurring patterns in word co-occurrence to generate clusters of keywords. These keyword lists are then interpreted by human analysts to assign themes. For example, Abram et al. used LDA to identify themes for nurse interviews on substance use field [10]. However, this still required manual review of model-generated keywords to identify meaningful themes. Supervised NLP techniques, such as supervised BERTopic [11], aim to identify themes that align with human predefined labels. However, thematic analysis is fundamentally an inductive process, where researchers typically begin with a small number (10-20) of transcripts to discover previously unknown themes. Because this task is to generate themes rather than apply existing ones, supervised NLP approaches are conceptually incompatible with thematic analysis. As a result, traditional NLP approaches either depend on human interpretation or struggle to adapt supervised frameworks to inductive discovery, making them less adaptable to the dynamic, context-rich narratives characteristic of qualitative healthcare research.

Recent advances in large language models (LLMs), such as GPT-4 [12], offer promising solutions to these limitations. LLMs can analyze long text and generate human-readable outputs in zero-shot or few-shots settings, only requiring instructions or a small number of labeled examples. This capability makes LLMs particularly well-suited for qualitative research contexts that lack annotated data or heavily rely on manual interpretation, effectively addressing key limitations of traditional NLP approaches. Prior studies have demonstrated the potential of LLMs in qualitative analysis. For example, Renard et al. suggested that LLMs can uncover hidden insights from patient interview transcripts and identify dominant themes [13]. Similarity, Mannstadt et al. demonstrated that LLMs can rapidly identify dominant themes from patient interview transcripts, serving as a helpful complement to human analysis [14]. Another study applied LLMs to identify themes about cancer patients' experiences, demonstrating that LLMs perform well in capturing structural, temporal, and logistical aspects of narratives [15]. However, because these instructions were broad, the outputs often defaulted to generic patterns and overlooked emotional nuance and contextual depth.

To address these gaps, we developed **Co**llaborative **T**heme **I**dentification Agents (CoTI) to support qualitative thematic analysis with LLMs. Although frameworks like Thematic-LM [16], TAMA [17], and Auto-TA [18] have pioneered the use of multi-agent systems for social media and clinical interview data, our framework is specifically designed to capture the objective-specific insights often overlooked by general-purpose agents. CoTI integrates three specialized agents: *Instructor* is responsible for producing tailored instruction prompts to capture objective-specific insights, including psychosocial, emotional, and contextual dimensions that are often overlooked by broad instructions, while *Thematizer* and *CodebookGenerator* are designed to reflect the two key phases of the analytical workflow, with *Thematizer* extracting supporting excerpts (*clues*) and identifying initial codes for each transcript and *CodebookGenerator* summarizing these codes with similar meanings across all transcripts into themes, as each transcript contained its own set of codes that often overlapped conceptually but varied in wording. While CoTI is capable of operating as a fully automated system, LLMs are often used alongside human researchers in practice, tasked with reviewing, refining or validating LLM-generated outputs. However, how human-AI collaboration can enhance thematic analysis quality remains insufficiently understood. To explore this, we embedded CoTI's *Thematizer* in a user-facing application that enables real-time interaction with human. This implementation provides an opportunity to examine whether combining LLMs with human involvement can improve the quality of thematic analysis in healthcare research contexts.

Overall, the aim of this study was to develop and evaluate a multi-agent LLM framework for improving the efficiency of qualitative thematic analysis and to examine whether human–AI collaboration can enhance the quality of thematic analysis in healthcare research.

## Methods

### Data Collection

We used 10 interview transcripts that our research team collected for our previous study [6], and conducted 2 more interviews for the current study. Specifically, we conducted a qualitative study using one-on-one, semi-structured interviews with older adults (age $\geq$ 65 years) who were hospitalized in the acute cardiac care units at Memorial Hermann Hospital at Texas Medical Center and excluded patients diagnosed with heart failure for the first time during the hospitalization, those unable to respond appropriately due to mental status changes, or those who declined to participate [6].

The interview guide focused on four key questions: (1) participants' perceptions of their heart medication intensity, (2) situations in which they would feel the medications excessive, (3) factors that would make medication management easier, and (4) their overall issues in medication management [6]. After conducting in-person interviews with each participant in the hospital, we transcribed the audio recordings via both professional transcription and OpenAI's Whisper model (small size) [19]. Transcripts were reviewed by interviewers to ensure accuracy.

In addition, the *GPT-4o-mini* model was released on July 18, 2024 [20], and the *QwQ-32B* model was released on March 5, 2025 [21], whereas our research team published the previous results about patients' perceptions of heart failure medications on February 4, 2025 [6]. Although there was a slight temporal overlap between the release of *QwQ-32B* and our earlier publication, we believed that all model development and data analyses had been completed prior to the release of *QwQ-32B*. Therefore, there was no possibility of data leakage or model memory.

### Thematic Analysis Settings

We employed four different settings to perform thematic analysis. First, two senior investigators independently extracted clues and identified initial codes for each transcript and developed the codebook with themes using inductive and deductive thematic analysis. Two additional independent senior investigators then reviewed all transcripts and refined themes [6]. These final themes served as the reference standard in our study. Second, four junior investigators (working independently from senior investigators) were assigned a subset of transcripts (see Supplementary Table S1) and independently reviewed the transcripts to identify initial codes manually. Third, our proposed CoTI framework was applied to automatically extract clues and identify initial codes for each transcript and grouped similar codes across all transcripts to generate themes. Fourth, the same junior investigators used the web-based version of CoTI (see Interactive User Design) to extract clues and identify initial codes for their assigned transcripts.

## CoTI Model

### Model Summary

CoTI implements its multi-agent design through three LLM agents to mimic the traditional thematic analysis phases proposed by Braun and Clarke in 2006 (Fig. 1) [22]. Specifically, CoTI operationalizes phases that are relatively structured and reproducible, including familiarization, coding, and theme review [22]. We adopted a multi-agent design because these steps involve distinct analytical functions. Separating these functions improves transparency and allows intermediate outputs to be reviewed more easily than a single LLM. Within this design, *Instructor*, implemented using the *QwQ-32B* reasoning model [21], generates high-quality, tailored instruction prompts that will guide *Thematizer*'s analysis toward capturing contextual depth, thereby supporting the familiarization phase. *Thematizer*, built on the *gpt-4o-mini* model [20], extracts clues, generates reasoning and identifies initial codes for each transcript, thereby reproducing the coding phase of the traditional thematic analysis while also allowing fast and efficient collaboration with human to refine its outputs. *CodebookGenerator*, also based on the *gpt-4o-mini* model, summarizes similar codes across all transcripts into themes, supporting the theme reviewing phase.

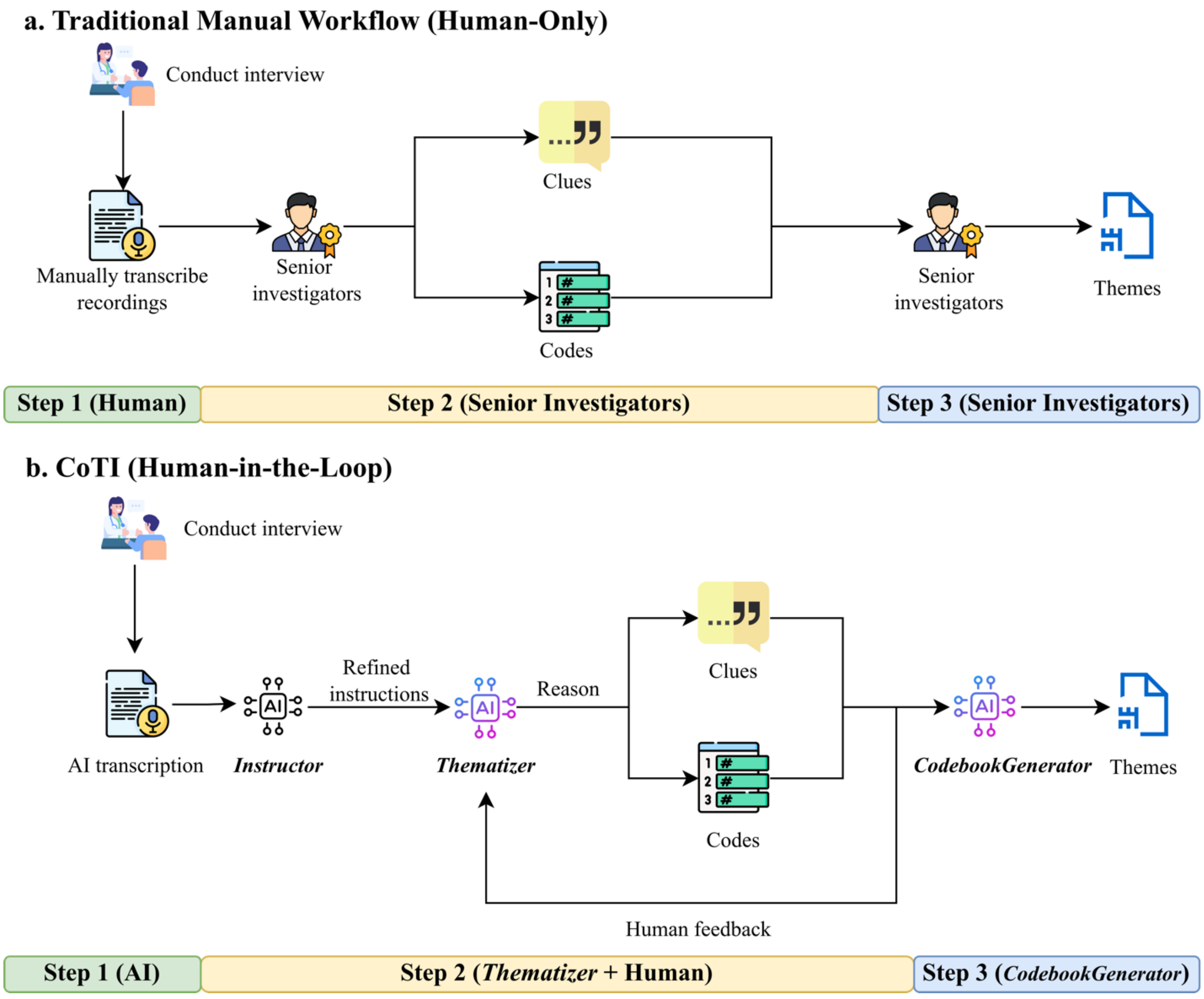

**Figure 1. Study Overview.** To explore treatment burden perceptions among older adults with heart failure, our study compares the traditional thematic analysis with our proposed CoTI framework. **a. Traditional Workflow:** The manual method consists of three main steps. First, researchers conduct semi-structured interviews and manually transcribe the audio recordings. Next, several senior investigators review each transcript to extract clues and identify initial codes. Finally, senior investigators generate themes based on clues and codes across all transcripts. This method is highly time-consuming and heavily reliant on expert labor. **b. CoTI Workflow:** In contrast, the CoTI framework introduces an human-AI collaborative workflow that improves scalability while preserving human involvement. First, interviews are transcribed using *Whisper*, an automatic speech recognition tool. Before thematic analysis, *Instructor* (based on a heavyweight reasoning-oriented LLM) iteratively improve the instruction prompts that will guide thematic analysis. Once finalized, these instructions are injected into *Thematizer* (based on a lightweight general-purpose LLM) to extract clues, generate reasoning, and identify initial codes. Human (e.g., junior investigators) can review these AI-generated outputs and provide feedback to refine them. Outputs across all transcripts are processed through *CodebookGenerator* (based on a lightweight general-purpose LLM) to generate themes.

## Instruction Generation Phase

**Overview.** In order to obtain high-quality instruction prompts for guiding thematic analysis while minimizing expert labor, we implemented an iterative instruction refinement process using *Instructor*. It began with the random selection of several transcripts, which were submitted to a reasoning model to identify initial codes. These AI-generated codes were not treated as the final codes for the transcripts but served as provisional references to guide instruction prompts development. *Instructor* took these AI-generated codes as inputs and progressively produced the refined clue and reasoning instruction prompts through four interconnected stages: *clue instruction*, *reasoning instruction*, *evaluation*, and *optimization*. Each stage built upon the previous one, ensuring a systematic progression toward high-quality clue and reasoning instruction prompts (Fig 2). Importantly, no senior investigators derived reference standard was provided during this process. The optimization procedure relied solely on the study objective, transcript content, and the provisional AI-generated codes.

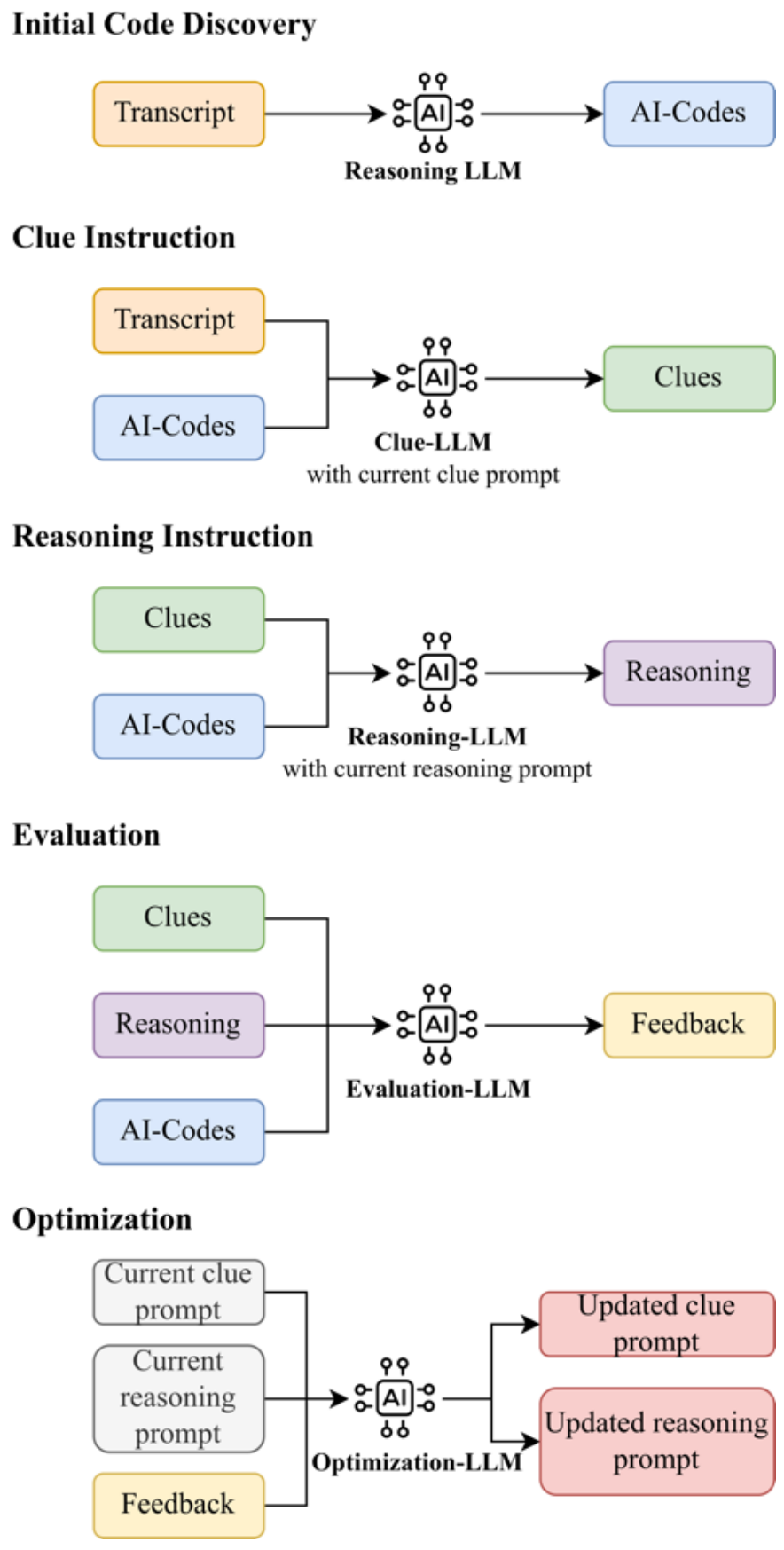

**Figure 2. Workflow of Instructor.** *Instructor* is a multi-agent framework designed to iteratively refine instruction prompts for *Thematizer* to perform thematic analysis. The process begins with a reasoning model (*QwQ-32B*) generating initial codes from several randomly selected transcripts. *Instructor* then launches a four-agent optimization cycle: ***clue-LLM***, ***reasoning-LLM***, ***evaluation-LLM***, and ***optimization-LLM***. First, *clue-LLM*, using the current clue prompt, takes each transcript and its corresponding AI-generated codes as input to generate clues for each code. Then, *reasoning-LLM*, guided by the current reasoning prompt, receives these clues and codes to generate reasoning statements. Next, *evaluation-LLM* takes every (clue, reasoning, code) triplet to produces feedback outlining issues with the current prompts and suggestions for improvement. Finally, *optimization-LLM* takes this feedback with the current clue and reasoning prompts to generate updated versions. These updated prompts replace the previous ones, initiating the next iteration of refinement.

**Initial Code Discovery.** To initiate the refinement process, we used the *QwQ-32B* reasoning model to identify initial codes in two randomly selected transcripts. The model was prompted to identify codes related to patients' perceptions of the treatment burden or intensity of heart failure medications (see prompts at Supplementary A.1).

**Clue Instruction.** The first stage of refinement focused on extracting clues, to support the interpretation of identified codes. For each training transcript paired with its corresponding AI-generated codes, we prompted a LLM agent (referred to as *Clue-LLM*, implemented using the *QwenQ-32B* reasoning model) with an initial instruction, "*List clues (i.e. key phrases, contextual information, semantic and emotional tones, temporal information, symptom descriptions) in the following patient-doctor dialogue that support each given identified topic.*"[26] (see prompts at Supplementary A.2). This instruction guided the model to find supporting excerpts for each code directly from the transcript.

*Clue-LLM* was specifically instructed to find clues in the form of direct quotes from transcripts, as these preserve the exact language and context, ensuring the original meaning. Unlike summaries or interpretations, direct quotes provide original and unchanged references, reducing bias and enhancing transparency. They also serve as reliable and traceable memory units, locating specific parts of the interview.

**Reasoning Instruction.** The second stage aimed to formalize the logical relationships between the extracted clues and their associated codes through structured reasoning. Using the clues generated by *Clue-LLM*, we prompted a second LLM agent (referred to as *Reasoning-LLM*, implemented using the *QwenQ-32B* reasoning model) with an initial instruction, "*Based on the given clues, generate the reasoning process that supports the identified topics.*"[26] (see prompts at Supplementary A.3). This enabled the model to generate structured reasoning statements that clarified how the provided clues supported each code, thereby enhancing the clarity and explainability.

**Evaluation.** The third stage involved assessing the quality of the generated clues and reasoning. Rather than collecting feedback for each individual (*clue, reasoning, code*) pair, we prompted a third LLM agent (referred to as *Evaluation-LLM*, implemented using the *QwenQ-32B* reasoning model) to identify common issues and offer overall suggestions for improving clue and reasoning instructions (see prompts at Supplementary A.4). This aggregated feedback addressed several limitations associated with individual feedback. Individual evaluations may result in inconsistent insights and overemphasize isolated patterns while overlooking systemic issues. Additionally, providing detailed feedback for each pair may increase complexity and reduce clarity in the evaluation process. By synthesizing feedback across multiple training examples, *Evaluation-LLM* was able to provide a more comprehensive assessment, enabling consistent and scalable improvements.

**Optimization.** The final stage focused on refining the clue and reasoning instructions based on feedback provided by *Evaluation-LLM*. During this step, a fourth LLM agent (referred to as *Optimization-LLM*) simultaneously improved the clue and reasoning instructions to address previously identified issues and enhance their overall quality. To mitigate the impact of potentially spurious feedback from the *Evaluation-LLM*, we prompted *Optimization-LLM* (implemented using the *QwenQ-32B* reasoning model) with the instruction "*The feedback may be noisy, identify what is important and what is correct.*" [23] (see prompts at Supplementary A.5). This prompt encouraged *Optimization-LLM* to apply critical thinking, allowing itself to prioritize relevant and accurate suggestions. The overall refinement process in the preprocessing phase was iterative, involving multiple cycles of clue and reasoning generation, evaluation, and optimization.

**Thematic Analysis Phase**

**Overview.** The thematic analysis phase aimed to apply the refined clue and reasoning instruction prompts to identify initial codes for all transcripts using *Thematizer*, and subsequently summarize these codes across all transcripts into a codebook with themes by *CodebookGenerator*.

**Codes Identification.** To enable rapid code identification for a given transcript, particularly in the setting where junior investigators' feedback is incorporated, we prompted *Thematizer* (implemented using the *GPT-4o-mini* model, which has faster inference speed compared with *QwQ-32B* model used in *Instructor*) with the refined clue and reasoning instructions obtained from the instruction generation phase (see prompts at Supplementary A.6). To capture a broader set of candidate codes, we repeated the code identification process three times and took the union of all codes generated across the three runs (see prompts at Supplementary A.7).

**Themes Generation.** Because each transcript had its own set of codes after *Thematizer*, many of which overlapped but differed in wording, we employed *CodebookGenerator* (implemented using the *GPT-4o-mini* model, for its faster inference speed) to group similar, semantically related, or duplicate codes across all transcripts into higher-level themes (see prompts at Supplementary A.8). Each resulting theme included a theme name, a description, the original codes included, and representative clues, which was suitable for human review.

### Human-AI Collaboration Web-based Application

We converted the CoTI framework into a user-friendly, web-based application to facilitate interaction between human (e.g., junior investigators) and our model (Fig 3). The application was built on *Thematizer*, which is responsible for extracting clues and identifying initial codes from individual transcripts. Because themes were generated through the analysis of all transcripts, and junior investigators in our study were each assigned only a subset of transcripts, *CodebookGenerator* was not converted into a human-collaboration module in our design. We selected a COVID-19 transcript as a demonstration example [24]. Within the application, users (e.g., junior investigators) are prompted to enter their Azure API credentials and upload a transcript. The uploaded transcript is displayed in a viewing panel on the right-hand side of the interface. Upon clicking the "Process Transcript" button, *Thematizer* automatically analyzes the transcript to extract clues and identify corresponding initial codes. The outputs are displayed for user review, allowing them to locate the relevant text segments in the transcript using keywords derived from the extracted clues (Figure 3a). If users are not satisfied with the outputs, they are offered two options: "Try Again", which reprocesses the transcript without feedback, or "Provide Feedback", which refines the model's outputs based on the user's input. This iterative feedback loop continues until the user indicates satisfaction by selecting "I'm Satisfied", at which point the final results are saved (Figure 3b).

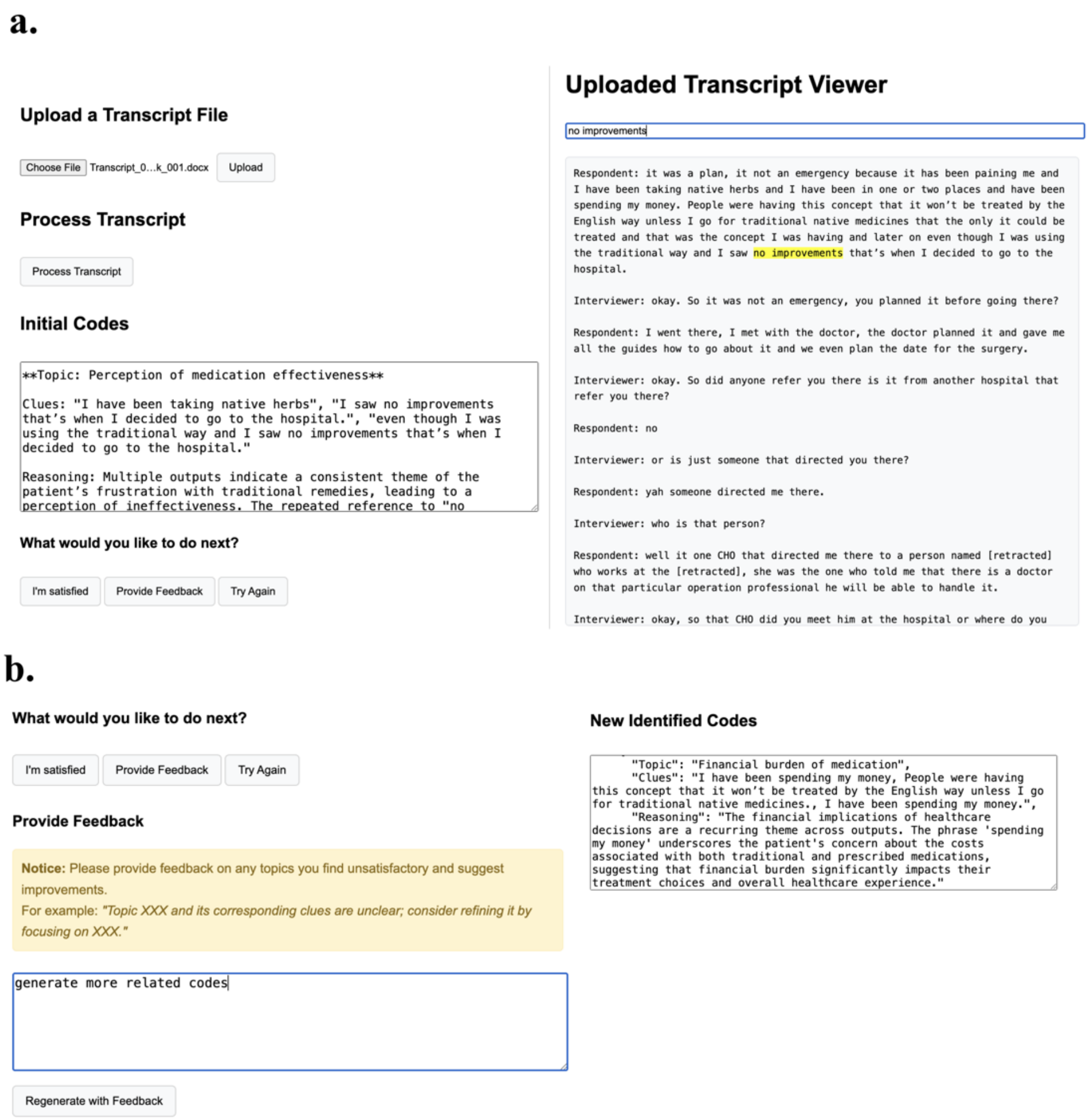

**Figure 3. Web-based APP. a.** After entering the Azure OpenAI API credentials and uploading a transcript, users can view the content on the right side of the interface. **b.** Once the initial model-generated response is provided, users are presented with three options: **“I’m satisfied”, “Try again”,** and **“Provide feedback”**. Selecting **“I’m satisfied”** indicates that the user accepts the response and concludes the code identification process. Choosing **“Try again”** prompts the model to reprocess the transcript and generate a new response. If **“Provide feedback”** is selected, the application displays text boxes where users can enter feedback. The model then incorporates this feedback to refine its previous response.

## Evaluation Against Senior Investigators

Because qualitative analysis is inherently interpretive rather than purely objective, thematic analysis cannot be considered as a simple pattern-recognition task. Accordingly, the purpose of our evaluation was not to determine whether model outputs were objectively corrected, but to assess how completely the model captures insights identified by senior investigators.

To achieve it, we evaluated our model's output (*clues*, *codes*, *themes*) against these provided by senior investigators, which served as the reference standard. We first compared model-extracted clues with these provided by senior investigators for each transcript. To quantify similarity, we used multiple metrics, including Jaccard similarity, precision, recall, and F1 score. Jaccard similarity measured the degree of overlap between model-extracted and senior-extracted clues. Precision reflected the proportion of correctly extracted clues among all extracted clues, while recall captured the proportion of reference clues that the model successfully extracted. Next, to evaluate code similarity, we calculated the cosine similarity between model-identified codes and those identified by senior investigators using embeddings generated from OpenAI's *text-embedding-3-small* model, which was independent of the models used in our main framework. Specifically, for each reference code, we identified the maximum cosine similarity with any model-generated code, averaged these values across all reference codes, and then averaged them across all transcripts. Finally, we assessed themes similarity by comparing the model-generated themes with the themes generated by senior investigators, using the same embedding-based cosine similarity method applied in the code evaluation.

**User Perception Survey**

To understand junior investigators' perceptions of CoTI's outputs and their experience, we implemented a survey guided by the QUEST framework [25] (see detailed questionnaire at Supplementary B). We had four independent junior investigators (medical school students who were not involved in the original theme generation project conducted by senior investigators) participating in this phase. Each junior investigator first reviewed their assigned transcripts to identify relevant codes manually, and then used the web-based version of CoTI application to perform the same task, after which they completed the survey.

**Baseline Models**

We established two types of baseline models for comparison. The first type included traditional unsupervised topic modeling methods such as Latent Dirichlet Allocation (LDA) [9], Top2Vec [26] and BERTopic [27]. These methods were selected because they are commonly used NLP approaches for identifying latent topics in textual data. However, they are not designed to perform the full qualitative thematic analysis workflow, such as extracting supporting excerpts and identifying initial codes for each transcript. Particularly, these baseline models take all transcripts as input and typically require preprocessing steps including data cleaning, tokenization, stopword removal, and lemmatization. Each baseline model outputs a set of latent themes represented as ranked lists of high-probability keywords. Since these keyword-based representations differ from human-interpretable themes, we developed a standardized evaluation framework for comparison. Specifically, we extracted the top 6 keywords from each baseline model's output as proxies for the generated themes. We then converted both keywords and senior investigators generated themes into embeddings and computed cosine similarity between them to assess theme quality.

The second baseline type (referred to as basic LLM) employed a reasoning-oriented model (*QwQ-32B*) without refined instructions for clue extraction and reasoning generation. Because reasoning-oriented LLM typically have longer inference times (*GPT-4o-mini* model required approximately one minute per transcript to complete three runs with aggregation, whereas a single run of *QwQ-32B* took approximately three minutes per transcript), we evaluated this baseline using a single run per transcript rather than the multiple aggregated runs used for CoTI. Unlike traditional models, this baseline was capable of generating clues, codes, and themes, allowing for direct comparison against senior investigators across all components.

### Ethics Approval

The interview study was conducted in accordance with the declaration of Helsinki and approved by UT IRB (HSC-MS-21-0874). This study has been approved by the Committee for the Protection of Human Subjects (the UTHSC-H IRB) under protocol HSC-SBMI-13-0549.

## Results

### Overview

We developed CoTI, a multi-agent human-AI collaborative framework designed to automate qualitative thematic analysis. Our goal was to efficiently generate high-quality outputs (*clues*, *codes*, and *themes*) that were similar to those produced by human experts (e.g., senior investigators). We evaluated CoTI's performance across three tasks: i) clue extraction (an intermediate output, i.e., supporting excerpts from transcripts), ii) codes identification (the primary output, i.e., codes derived from clues) within each transcript, and iii) theme generation (across all transcripts, summarizing all codes into a codebook with themes). Our experiments showed that clues, codes, and themes that identified by CoTI were more similar to those of senior investigators than were the outputs of traditional NLP models, basic LLM, or human researchers with lower levels of experience (e.g. junior investigators). Moreover, collaboration between CoTI and junior investigators did not lead to outputs that were more similar to those of the senior investigator than CoTI alone.

### Interview Data Collection and Patient Characteristics

As a case study, we conducted interviews with 12 heart failure patients from Memorial Hermann Hospital at the Texas Medical Center in Houston, Texas, to explore their perceptions of challenges in using heart failure medications [6]. Among the 12 participants, 8 (66.67%) were female, with a mean age of 74.75 years (SD = 8.21). 5 participants (41.67%) were White, and 5 (41.67%) were African American. Detailed demographic and clinical information for each participant is presented in Table 1.

**Table 1**. Participant Demographic and Clinical Characteristics

| Age | Sex | HF type | Comorbid conditions | Number of prescribed medications |
|---|---|---|---|---|
| 71 | female | HFpEF | 9 (atrial fibrillation, anemia of chronic disease, chronic kidney disease, diabetes mellitus, hypertension, hyperlipidemia, morbid obesity, obstructive sleep apnea) | 16 |
| 82 | female | HFrEF | 3 (breast cancer, hyperlipidemia, hypertension | 20 |
| 67 | female | HFpEF | 5 (stroke, pulmonary embolism, atrial fibrillation, hypertension) | 7 |
| 74 | male | HFpEF | 4 (atrial fibrillation, chronic obstructive pulmonary disease, diabetes mellitus, hypertension) | 7 |
| 69 | female | HFrEF | 6 (atrial fibrillation, alcohol abuse, anemia of chronic disease, hypertension, hypothyroidism and mild liver disease) | 6 |
| 70 | female | HFpEF | 8 (coronary artery disease, chronic obstructive pulmonary disease, stroke, diabetes mellitus, hypertension, hypothyroidism, peripheral arterial disease, obstructive sleep apnea) | 15 |
| 66 | female | HFpEF | 9 (chronic obstructive pulmonary disease, coronary artery disease, diabetes mellitus, gastroesophageal reflux disease, hypertension, hyperlipidemia, hypothyroidism, morbid obesity, obstructive sleep apnea) | 14 |
| 75 | male | HFrEF | 6 (coronary artery disease, atrial fibrillation, end-stage-renal disease, hypertension, hyperlipidemia, hypothyroidism) | 13 |
| 85 | male | Unknown | 7 (atrial fibrillation, diabetes mellitus, hyperlipidemia, hypertension, hypothyroidism, major depression, peptic ulcer disease) | 11 |
| 88 | female | HFpEF | 7 (coronary artery disease, chronic obstructive pulmonary disease, cerebral venous sinus thrombosis, hypertension, atrial fibrillation, chronic kidney disease, obstructive sleep apnea) | 14 |
| 85 | female | HFpEF | 4 (mitral regurgitation, hypertension, chronic kidney disease, monoclonal gammopathy of undermined significance) | 8 |
| 76 | male | HFrEF | 5 (atrial fibrillation, hypertension, chronic kidney disease, stroke, hyperlipidemia) | 11 |

**HFpEF**: heart failure with preserved ejection fraction; **HFrEF**: heart failure with reduced ejection fraction.

### CoTI Produced Codebook Similar to Senior Investigators' Codebook

To evaluate whether CoTI could perform thematic analysis similarly to senior investigators, both senior investigators and CoTI extracted clues, identified codes, and developed a codebook with themes, respectively. We then compared the similarity between them.

We first utilized *Instructor* to refine instruction prompts that will guide *Thematizer* (See initial code discovery output at Supplementary C.1). As shown in Table 2, the instruction prompt to extract clues has evolved from an initial, general request for key phrases into a more structured instruction emphasizing direct quotes, contextual completeness, exclusive code assignment, and causal clarity. Similarly, the instruction prompt to generate reasoning advanced to require code-specific, stepwise causal chains, and precise language grounded solely in provided clues.

**Table 2**. Optimized instruction prompts by *Instructor*

| | To extract clues | To generate reasoning |
|---|---|---|
| Before optimization | List clues (i.e. key phrases, contextual information, semantic and emotional tones, temporal information, symptom descriptions) in the following patient- | Based on the given clues, generate the reasoning process that supports the identified topics. |
| | doctor dialogue that support each given identified topic. | |
| After optimization | Extract **direct quotes** from the dialogue that explicitly support each topic, ensuring:<br>1. **Contextual Completeness & Causal Links**: Include specific details that clarify mechanisms or outcomes (e.g., *"After doubling the dose, my BP remained at 160/100 despite the doctor's adjustment"* instead of *"dose increase failed"*). Specify quantitative data, professional feedback, or patient-reported outcomes to strengthen causal relationships.<br>2. **Exclusive Topic Assignment**: Assign each quote to only one topic unless it explicitly addresses multiple themes *simultaneously* (e.g., *"My fixed income can't cover my 12 pills daily"* links both financial burden and polypharmacy). Avoid cross-topic bleeding (e.g., *"too many pills"* for cost vs. adherence).<br>3. **Clarity in Reuse**: For quotes used across topics (e.g., religious coping statements), append contextual phrases to clarify relevance (e.g., *"I put everything in the Lord's hands [to cope with stress]"* for psychological themes vs. *"…to accept my medication burden"* for adherence topics).<br>4. **Causal Precision**: Prioritize quotes establishing explicit cause-effect chains (e.g., *"The rash from the new pill made me stop taking it"* instead of *"I stopped the pill"*). Specify whether effects are patient-reported, caregiver-observed, or clinically measured.<br>5. **Discrepancy Framing**: When perspectives conflict, frame quotes within the topic's context (e.g., *"Patient says 'I take all meds,' but caregiver notes 'she skips 3 pills weekly'"* under adherence | For each topic, construct a logical chain connecting clues to the topic by:<br>1. **Numbered Stepwise Causality**: Break down causal pathways into explicit, sequential steps (e.g., *"Step 1: Eliquis caused bleeding → Step 2: Fear of overmedication → Step 3: Reduced adherence → Step 4: Uncontrolled condition"*).<br>2. **Mechanism & Behavioral Impact**: Specify *how* each clue leads to outcomes, including patient behavior changes (e.g., *"Step 1: High pill count → Step 2: Cognitive overload → Step 3: Missed doses → Step 4: Worsened polypharmacy burden"*; *"Step 1: Financial strain → Step 2: Delayed ER visits → Step 3: Complication escalation"*).<br>3. **Avoid Assumptions**: Explicitly map clues to outcomes using only provided data (e.g., *"Step 1: Dose escalation caused nausea → Step 2: Nausea reduced medication intake → Step 3: Suboptimal BP control"* instead of implying indirect links).<br>4. **Address Contradictions**: Explain discrepancies as causal factors (e.g., *"Step 1: Patient denies non-adherence → Step 2: Caregiver notes missed doses → Step 3: Conflicting narratives → Step 4: Potential for unmanaged symptoms"*).<br>5. **Distinct Factor Differentiation**: Separate overlapping effects (e.g., *"Step 1: High dosage → Step 2: Nausea → Step 3: Reduced adherence"* vs. |
| | challenges).<br>Ensure quotes are concise but include sufficient detail to avoid ambiguity and support rigorous reasoning. | *"Step 1: Drug interactions → Step 2: Dizziness → Step 3: Fall risk"*).<br>6. **Actionable Language**: Use precise terms like *"triggers,"* *"results in,"* or *"directly causes"* to replace vague phrasing.<br>Ensure reasoning is topic-specific, free of redundancy, and grounded solely in provided clues. |

Then, we applied *Thematizer* with these refined instruction prompts to extract clues and identify initial codes from all (12) transcripts. We calculated Jaccard similarity, precision, recall, and F1 score between clues extracted by CoTI and those extracted by senior investigators to evaluate clue similarity. We also computed cosine similarity between initial codes generated by CoTI and those identified by senior investigators to measure code similarity. As shown in Table 3 and Table 4, our model produced clues and codes more similar to those of senior investigators than did the basic *QwQ-32B* LLM across most evaluation metrics. Specifically, relatively greater gains were observed in Jaccard score (+7.8%) and precision (+10.9%), with modest gains in F1 score (+5.4%) and cosine similarity (+4.9%). The response time to extract clues and identify codes was approximately 1 minute per transcript for CoTI and 3 minutes per transcript for the basic *QwQ-32B* model.

After generating code for each transcript, we employed *CodebookGenerator* to summarize codes across all transcripts into a structured codebook with themes (See outputs at Supplementary C.2). We calculated the cosine similarity between the CoTI-generated and senior-generated themes to evaluate theme similarity. As illustrated in Table 4, CoTI generated themes that was more similar to the senior investigator's than those produced by traditional NLP topic modeling methods (+ 58.82%) or the basic *QwQ-32B* LLM (+22.24%). The codebook generation time was approximately 10 seconds for CoTI and 2 minutes for the basic *QwQ-32B* model. Table 5 shows the overlap between the CoTI- and senior-generated themes. CoTI successfully captured several major themes identified by senior investigators, including adverse drug effects, psychological distress, burden from the number of medications, and burden from the cost of medication.

**Table 3**. Evaluation of clue similarity on the heart failure interview transcripts.

| | Clue Extraction | |
|---|---|---|
| | Basic *QwQ-32B* LLM | CoTI |
| Jaccard Similarity | 0.374 | 0.403 (+7.75%) |
| Precision | 0.496 | 0.550 (+10.87%) |
| Recall | 0.635 | 0.630 (-0.79%) |
| F1 Score | 0.540 | 0.569 (+5.37%) |

**Table 4.** Evaluation of codes and themes similarity on the heart failure interview transcripts.

| | Code Identification | | Theme Generation | | | | |
|---|---|---|---|---|---|---|---|
| | Basic *QwQ-32B* LLM | CoTI | LDA | Top2Vec | BerTopic | Basic *QwQ-32B* LLM | CoTI |
| Cosine Similarity | 0.411 | 0.431 (+4.87 %) | 0.391 | 0.377 | 0.266 | 0.508 | **0.621** |

**Table 5.** Comparison of codebook with themes developed by senior investigators and CoTI (without junior investigators) on the heart failure interview transcripts.

| Senior Investigator | Overlap | CoTI |
|---|---|---|
| Problems in Logistics | Adverse Drug Effects | Medication Adherence and Management Challenges |
| Impact from the patient-doctor relations | Psychological distress | Desire for Simplified Medication Regimen |
| | Burden from the number of medications | Perceived Effectiveness of Medications |
| | Burden from the cost of medications | Patient-Doctor Relationship and Communication |

### CoTI Alone Was More Similar to Senior Investigators Than Junior Investigators or CoTI–Junior Collaboration

After verifying that CoTI can generate themes more similar to that of senior investigators than other NLP models or the basic LLM, we conducted an exploratory human-AI collaboration experiment to examine whether a human (particularly one with low expertise, such as a junior investigator) could further refine CoTI-generated outputs. The purpose of this experiment was not to compare CoTI with junior investigators performing the full thematic analysis workflow from scratch. Instead, because CoTI is designed to improve efficiency of thematic analysis by generating clues, initial codes, and themes for human review, we evaluated whether junior investigators could use CoTI-generated outputs as a starting point and refine them.

We compared three settings (Table 6): (1) Junior Investigator alone, where junior investigators independently identified codes for a subset of assigned transcripts; (2) CoTI alone, where our model extracted clues, identified codes, and generated themes without junior investigators' feedback for all transcripts; and (3) CoTI + Junior Investigator, where junior investigators used CoTI application to review model-generated clues and codes, and provided feedback to CoTI to refine its outputs for each assigned transcript. In all settings, clues, codes and themes provided by senior investigators served as the reference standard. Since we considered codes as the final analytic output of each transcript, junior investigators were only responsible for manually identifying codes and did not perform clue extraction. Additionally, because the themes can be generated only after considering codes across all transcripts and junior investigators were assigned only a subset of transcripts, they could not construct a complete codebook with themes. Therefore, the evaluation focused on clue extraction for the CoTI alone and CoTI + Junior Investigator settings, and on code identification for all three settings.

**Table 6.** Four different settings to perform thematic analysis on the heart failure interview transcripts.

| Setting | # of Interview transcripts to process | Extract clues | Identify codes | Generate themes |
|---|---|---|---|---|
| Junior investigator alone | A few (7) out of 12 | No | Yes | No |
| CoTI alone | All | Yes | Yes | Yes |
| CoTI + junior investigator | A few (7) out of 12 | Yes | Yes | No |
| Senior investigators only (reference standard) | All | Yes | Yes | Yes |

To evaluate whether junior investigators' feedback could help CoTI to extract clues that are more similar to the senior investigators, we compared clue similarity between CoTI alone and the senior investigators, and between CoTI with junior investigators' feedback and the senior investigators. Among the evaluation metrics, we prioritized recall as the most meaningful evaluation metric since higher recall indicates that the model successfully retrieves a large proportion of relevant clues that identified by senior investigators, which is critical for ensuring that subsequent code identification is grounded in a sufficiently rich evidence base. Although precision reflects the accuracy of extracted clues, occasional inclusion of clues not identified by senior investigators may be less damaging because qualitative thematic analysis is inherently subjective, additional identified results may also be valid even if they are not present in the reference standard. As shown in Figure 4a, CoTI with junior investigators' feedback resulted in small or modest recall improvements in many transcripts. While precision was not the primary metric, it offers complementary insight into the correctness of extracted clues. As shown in Figure 4b, CoTI with junior investigators' feedback generally yielded lower precision compared to the CoTI alone. Even in the few cases where precision improved after junior investigators' feedback, such as transcript 3 and 7, the gains were marginal. These results suggest that CoTI alone already provides strong clue similarity, and junior investigators' feedback brings limited or even negative impact on improving clue similarity with senior investigators. One possible hypothesis is that the gap between CoTI and senior investigators may involve domain-specific understanding that junior investigators also lack. As a result, junior investigators' feedback may not sufficiently refine the model's outputs. Another possible hypothesis is that junior investigators may have exhibited automation bias, a tendency to rely on outputs from CoTI, and this overreliance could have reduced their critical engagement with the model's outputs.

We continued evaluating the impact of junior investigator's feedback for the task of code identification. As shown in Figure 4c, codes identified by both CoTI alone and CoTI with junior investigators' feedback achieved higher similarity to senior investigators than codes manually identified by junior investigators in many cases. Notably, for some cases, incorporating junior investigators' feedback to CoTI led to identified codes that were more similar to those of senior investigators, as seen in transcripts 2, 3, 4, and 8, where CoTI alone achieved moderate theme similarity scores (approximately between 0.35 and 0.40). However, when CoTI alone already achieved strong code similarity to senior investigators (cosine similarity $\geq 0.45$), incorporating junior investigators' feedback tended to reduce that similarity, as observed in transcript 5, 6, and 10. Compared to the clue extraction task, junior investigators' feedback appeared more helpful in supporting the code identification task, suggesting junior investigators may be more adept at higher-level interpretation than at intermediate clue extraction.

The marginal benefit of junior investigator's feedback to CoTI might be due to junior investigators' perception on AI. To explore this possibility, we collected their perceptions of CoTI. As illustrated in Figure 5a, subjective ratings were generally high across all junior investigators, indicating strong approval of CoTI's accuracy, relevance, trustworthiness, and overall satisfaction. These favorable perceptions suggest that junior investigators may have felt less need to critically revise or question CoTI's outputs, thereby reducing the additive value of AI-human collaboration. However, we noticed several exceptions. Investigator D rated CoTI as missing codes in every assigned transcript, resulting in the code comprehensiveness score of 0%. This is particularly striking given that D simultaneously gave perfect trust (100%) and top scores across most other dimensions. To better understand this contradiction, we analyzed D’s performance in the manual code identification task. D achieved the code similarity of 0.41, higher than Investigator A and C, though slightly lower than CoTI alone (code similarity = 0.45). Additionally, D identified a total of 47 codes across assigned interviews, compared to 41 themes identified by CoTI. These results suggested that Investigator D was particularly context-sensitive and may have maintained a high bar for code completeness. Although D trusted CoTI’s identified codes, D likely expected CoTI not only to identify obvious codes but also to capture more implicit ones. In addition, collaboration between D and CoTI led to small gains in clue similarity compared to CoTI alone, which indicated that even for a higher standard evaluator, collaboration with AI offered marginal but observable benefits. Another notable exception was Investigator A, whose performance in code identification was comparatively weaker than investigators B and D. Moreover, CoTI alone achieved higher code similarity than A’s individual performance, and collaboration between A and CoTI improved clue similarity compared to CoTI alone. Nevertheless, Investigator A reported lower trust in CoTI compared with other investigators, indicating that A's confidence in AI remained low even when CoTI outperformed A's own performance and the collaboration yielded measurable benefits.

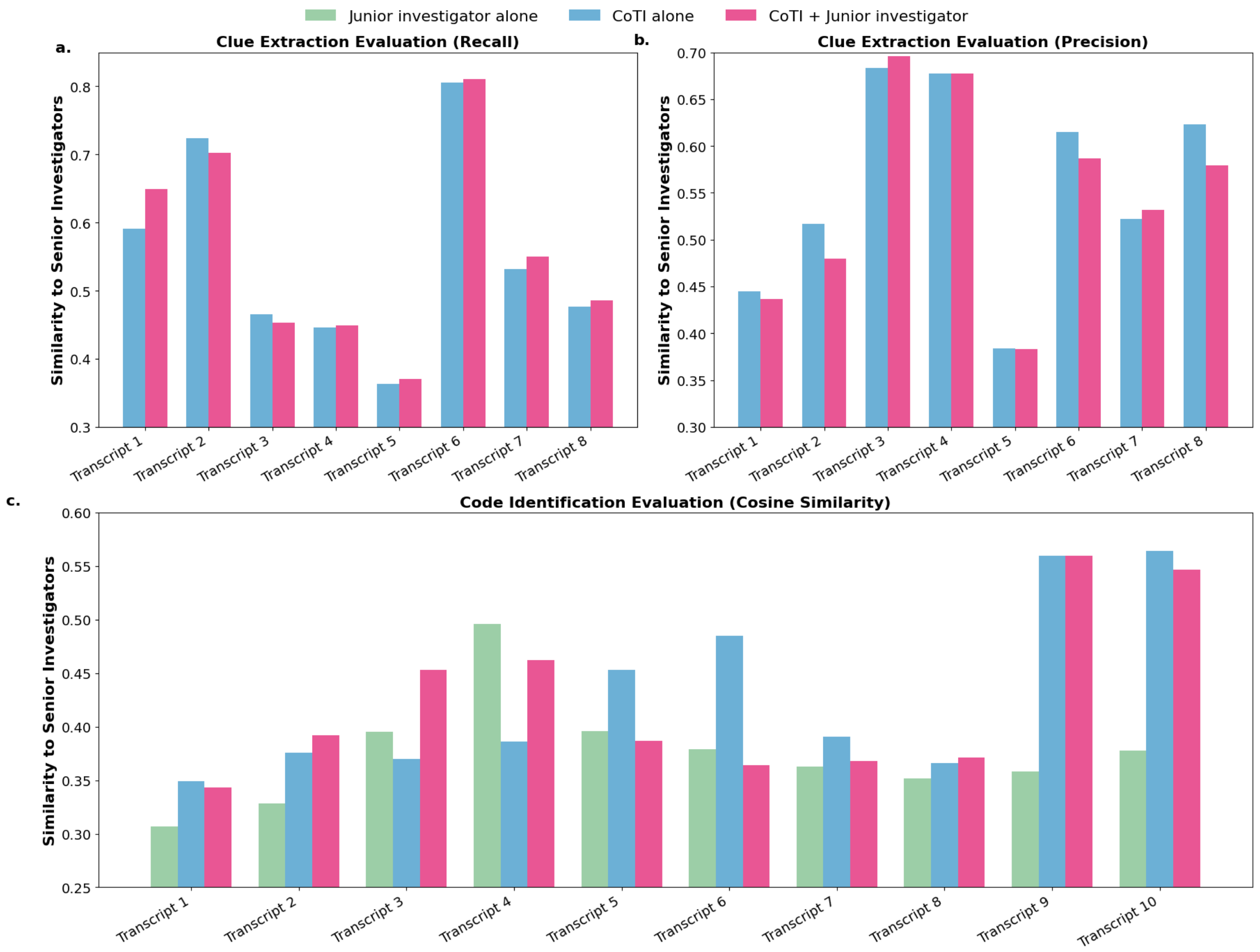

**Figure 4. Evaluation of clue extraction and code identification across interviews. a.** Recall of extracted clues for CoTI-only and CoTI + Human settings, evaluated against clues identified by senior investigators. **b.** Precision of extracted clues for CoTI-only and CoTI + Human settings, evaluated against clues identified by senior investigators. **c.** Cosine similarity between codes generated by Human-only, CoTI-only, or CoTI + Human, compared to senior investigators identified codes (reference standard).

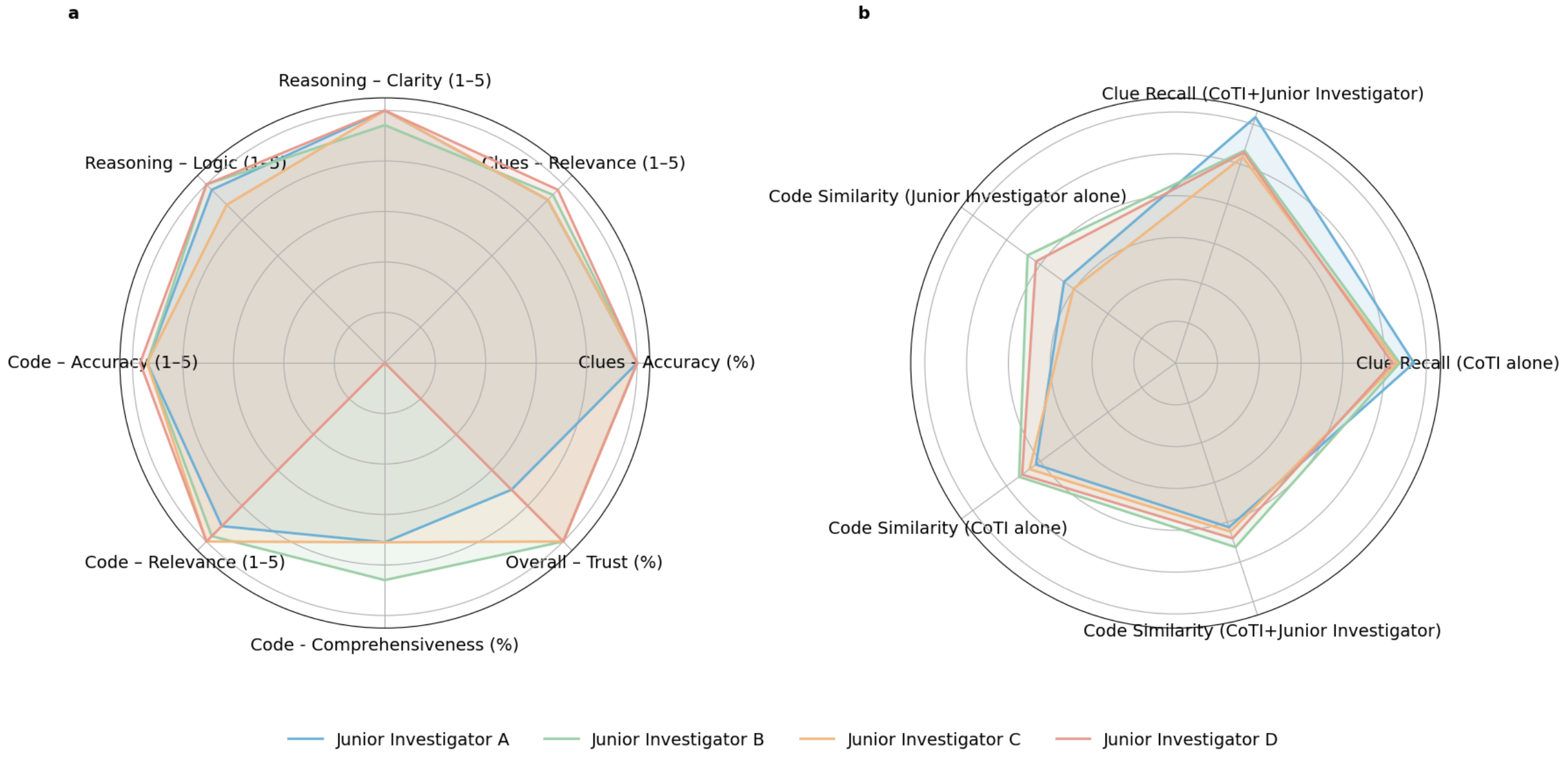

**Figure 5. Subjective Survey and Objective Evaluations by Junior Investigators. a.** This spider chart displays subjective perceptions from four junior investigators (A–D) regarding CoTI. Subjective measures include ratings of clue accuracy and relevance, reasoning clarity and logic, code accuracy, relevance, and comprehensiveness, as well as trust and overall satisfaction. **b.** This spider chart displays objective performance by junior investigators and CoTI, including clue recall (CoTI alone), clue recall (CoTI+Junior Investigator), code cosine similarity (Junior Investigator alone), code cosine similarity (CoTI alone), and code cosine similarity (CoTI+Junior Investigator).

## Ablation Study Highlights Component Contributions to CoTI Performance

Aiming to understand the contributions of each component within our model, we conducted an ablation study (Table 7). The study began with a GPT-4o-mini model, which serves as the baseline. In this setting, GPT-4o-mini performed thematic analysis only one time for each transcript (total 12). Introducing *Instructor*, an agent designed to iteratively refine instruction prompts, improved the similarity of extracted clues (Jaccard similarity: +14.3%, precision: 24.3%, recall: +2.7%, F1 score: 11.0%). These results highlighted the critical role of high-quality, tailored instruction prompts in improving CoTI's ability to extract clues that were more similar to those extracted by senior investigators. Subsequently, we applied a multi-run aggregation strategy, which repeated the code identification including clue extraction three times and retained the union of the resulting outputs. This strategy further improved performance, yielding relatively greater gains in clue similarity (Jaccard similarity: +17.5%, recall: +18.7%, F1 score: 12.5%) with modest improvements in code similarity (cosine similarity: +0.9%). These findings indicated that multiple code identification passes allowed our model to capture a broader range of codes and supporting clues, including these that may be missed in a single run due to the randomness in LLMs outputs.

**Table 7.** Overview of CoTI components and additive contributions to the performance.

| | **Clue Extraction** | | | | **Code Identification** | **Response Time for each transcript** |
|---|---|---|---|---|---|---|
| | Jaccard Similarity | Precision | Recall | F1 score | Cosine Similarity | Time |
| gpt-4o-mini | 0.300 | 0.441 | 0.517 | 0.456 | **0.433** | ~8 seconds |
| gpt-4o-mini with *Instructor* | 0.343 (+14.33%) | 0.548 (24.26%) | 0.531 (+2.71%) | 0.506 (+10.96%) | 0.427 (-1.39%) | ~20 seconds |
| CoTI (gpt-4o-mini with *Instructor* and multi-run aggregation) | **0.403** (+17.49%) | **0.550** (+0.36%) | **0.630** (+18.64%) | **0.569** (+12.45%) | 0.431 (+0.94%) | ~1 minute |

**Generalizability of CoTI on External Datasets**

To further evaluate the generalizability of CoTI, we applied our framework to several different healthcare qualitative datasets, including the health system's response to COVID-19 in Sierra Leone, which comprised 21 interview transcripts [24], and the family carers' strategies when a family member with dementia was agitated, which comprised 18 interview transcripts [28]. Table S2 shows the computational evaluations among these external datasets. Table S3-S4 shows overlaps between the themes generated by our model and by experts. Across these external datasets,

CoTI achieved higher clue similarity than theme similarity, indicating that it identified relevant supporting excerpts more effectively than it reproduced expert-level themes. This suggests that human researchers review remains necessary to interpret and validate generated themes.

## Discussion

Our study demonstrates the efficiency of leveraging multi-agent LLMs to support thematic analysis in qualitative research. By integrating prompt engineering with multi-agent collaboration, our proposed framework successfully extracted supporting excerpts, identified initial codes, and developed a codebook with themes through an automated process. Beyond its application in heart failure qualitative research, our framework has demonstrated its efficiency in other domains, such as COVID-19-related and Alzheimer-related studies. In addition, our developed CoTI interactive application allows human researchers to rapidly review model-generated outputs, locate key transcript sections by searching keywords derived from model-extracted supporting excerpts, and provide feedback to refine the outputs. This interactive design facilitates human-AI collaboration and may help researchers more efficiently gain insights into patients' experiences, thereby facilitating qualitative research. However, our preliminary experiment found that collaboration between CoTI and junior investigators yielded marginal improvements over CoTI alone in the heart failure study.

The computational evaluations showed that the themes generated by CoTI alone were more similar to those identified by senior investigators than those generated by junior investigators working independently or baseline models. To further examine the quality of these themes, we incorporated an expert qualitative appraisal as a complementary assessment. The expert identified overlap among the themes “medication burden”, “psychological and emotional impact of medications”, and “side effects and their perception”. In particular, the expert considered “side effects and their perception” unnecessary as a separate theme because these experiences can be integrated within either medication burden or the psychological and emotional impact of medications. Similarly, the expert noted that “desire for simplified medication regimen” can be integrated into “medication adherence and management challenges”. These observations suggested that CoTI may separate related experiences into multiple themes, resulting in thematic distinctiveness that are less clearly differentiated. This qualitative appraisal aligns with our computational evaluation, which indicated relatively modest theme distinctiveness (0.473).

Next, we compared CoTI-generated themes with the reference themes developed by senior investigators to identify where CoTI missed or partially distorted the reference interpretations. One missed example was the reference theme "problems in logistics". In senior investigators generated interpretations, this theme referred not only to medication taking behavior but also to broader practical barriers, including transportation difficulties, obtaining medications, coordinating refills, identifying the correct pharmacy, and relying on family members or caregivers to complete multiple steps in the medication process. CoTI did not generate an equivalent theme. Instead, it generated the theme “medication adherence and management challenges”, which focused more narrowly on patients’ difficulties with medication adherence, including forgetfulness, confusion, overdosing, and the need for structured support systems. One distorted example involved the reference theme "impact from patient-doctor relations". The senior investigators generated theme captured a complex pattern that included both trust and dependency on physicians, as well as skepticism toward pharmaceutical companies and insurance systems. CoTI generated a related theme “patient-doctor relationship and communication”, but this theme emphasized communication gaps and patient education rather than the broader trust–mistrust tension captured in the reference interpretations. Another partially distorted example was CoTI’s theme “desire for simplified medication regimen”. Although the supporting excerpts for this theme overlapped with those extracted by senior investigators, the senior investigators interpreted these excerpts as part of the broader theme “burden from the number of medications”. In contrast, CoTI considered patients’ preferences for a simplified or reduced medication regimen as a separate theme, rather than integrating them into medication burden. CoTI also generated a theme “perceived effectiveness of medications”, which was plausible but not supported by the reference interpretations. Although some patients discussed whether their medications were working, the senior investigators generated themes did not identify it as a separate theme. This suggested that CoTI may sometimes identify some less important patterns as distinct themes even when human researchers interpret it as secondary or insufficiently central to the overall research question. Together, these examples suggest that while CoTI can identify relevant semantic content, it may sometimes reorganize these patterns into themes that are more general, more familiar, or less central to the research question than those developed by human researchers. These findings indicate that CoTI can serve as an efficient tool to support thematic analysis, but its outputs still require careful human review and interpretation. This conclusion is consistent with Mannstadt et al., who emphasized that LLMs should complement rather than replace human analysis, particularly when identifying structural themes.[14]

Our study also provides preliminary insights into the potential role of CoTI in supporting human–AI collaboration in thematic analysis. First, CoTI alone produced codes more similar to senior investigators than junior investigators. This highlights CoTI’s potential as a reliable tool for thematic analysis, particularly when senior investigators are unavailable. Second, collaboration with junior investigators did not consistently improve clue or code similarity to senior investigators. While human-AI collaboration is often assumed to enhance AI outputs, our results show that the effectiveness is marginal. Specifically, when CoTI alone code similarity was moderate to senior investigators, junior investigator’s feedback could help refine codes and enhance similarity. However, when CoTI alone already performed strongly, such feedback often degraded code similarity. In addition, CoTI alone extracted clues that often achieved higher recall and precision, however, incorporating junior investigators’ feedback sometimes making clue similarity decrease. Survey analyses indicated that the marginal benefits of collaboration may stem from junior investigators' overreliance on CoTI's outputs, which could cause automation bias and reduce their independent critical thinking.

Our study has several limitations. First, the human-AI collaboration was exploratory. We recruited only four junior investigators, who analyzed selected subsets of transcripts and did not complete the full thematic analysis workflow, including excerpt extraction, code identification, and theme generation through consensus or adjudication process. Therefore, our findings only provided preliminary evidence that CoTI-generated outputs were more similar to the senior investigators generated reference standard than those generated by the participating junior investigators. Future studies should evaluate both CoTI alone and CoTI-assisted human–AI collaboration against rigorous human-led thematic analysis workflows involving multiple trained coders, consensus coding, and adjudication, thereby providing a more rigorous assessment of CoTI's effectiveness and the potential benefits of human–AI collaboration in thematic analysis. Second, the human-AI collaboration was conducted on a single case study of heart failure interviews, which limits the generalizability of our findings on human-AI collaboration in thematic analysis. Further validation across other research contexts is needed to establish broader generalizability. Moreover, the potential impact of incorporating feedback from senior investigators into CoTI was not evaluated, which may limit understanding of how senior investigators input could further enhance our model performance. Another limitation is that our interpretation of automation bias was based on a small exploratory survey of junior investigators' perceptions of CoTI. Although the survey results suggested that the high trust of CoTI may reduce junior investigators' tendency to critically revise AI-generated outputs, these findings remain preliminary because of the small sample size and reliance on self-reported perceptions. Moreover, alternative explanations cannot be excluded, such as limited qualitative analysis experience among junior investigators, lower quality of junior investigators' feedback. Therefore, we cannot consider automation bias or other factors as the definitive explanations for the limited additive benefit of junior investigators' involvements.

In all, CoTI provides preliminary evidence that multi-agent LLMs can replicate key phases of qualitative thematic analysis while maintaining human involvement through review and refinement. In addition, our study suggests that the value of LLMs in qualitative research lies not in replacing researchers, but in expanding their capacity to analyze larger and more complex datasets efficiently while maintaining interpretive oversight. Future research should focus not only on improving model performance, but also on designing collaborative workflows that preserve reflexivity and critical interpretation.

**Acknowledgements**

The authors used generative artificial intelligence (ChatGPT) to assist with manuscript language refinement. All AI-assisted text was reviewed, edited, and verified by the authors, who take responsibility for the accuracy and integrity.

**Funding Statement**

This work was supported in part by National Institute of Health (NIH) under award number R01AG082721 and R01AG084637.

**Data Availability**

The interview transcripts about heart failure patients cannot be made publicly available due to privacy concerns.
For interview transcripts about COVID-19, please see: https://osf.io/qc3z8/overview [24]
For interview transcripts about Alzheimer-related, please see:
https://data.mendeley.com/datasets/s8wtptnhyc/1 [28]

**Code Availability**

The code used in this study is available at https://github.com/QidiXu96/CoTI-MultiAgent-Theme-Analysis

**Author Contributions**

Concept and design: QX and YK. Data Collection: NA and MK. Model development: QX and YK. Senior investigator: NA. Junior investigators: GG, AC, DH and AW. Interpretation: QX and YK. Draft manuscript: QX, MK, and YK. All authors contributed to editing the manuscript, approved the final manuscript, and accepted the responsibility to submit it for publication.

**Conflicts of Interest**

Min Ji Kwak received a consulting fee from Novo Nordisk. Otherwise, no potential conflict of interest relevant to this article was reported.

**Abbreviations**

**HFpEF**: heart failure with preserved ejection fraction
**HFrEF**: heart failure with reduced ejection fraction
**LLM**: large language model
**NLP**: natural language processing
**LDA**: latent dirichlet allocation

**Supplementary**

A. **Prompts**
1. **Prompt for initial code discovery**

**System:**
### Background ###
Heart failure patients' perception of medication intensity is a complex experience influenced by factors beyond dosage, including side effects, treatment burden, psychological impact, quality of life, and cost. This perception is shaped by patient characteristics such as age, disease severity, comorbidities, gender, and socioeconomic status. Understanding these perceptions, along with patients' beliefs about their medications, is crucial for improving adherence and tailoring treatment strategies to enhance their quality of life. Ultimately, a patient-centered approach that involves effective communication and shared decision-making is essential in managing heart failure medication regimens.

You are a qualitative research expert tasked with identifying topics of patient-doctor dialogues. These patients are diagnosed with heart failure. You are trying to understand patient's perception of the intensity of heart failure medications.

Your task:
- Identify important topics for the given dialogue (there may be more than one).
- For each identified topic, provide clues and reasoning to explain the connection.

Clues must:
- Be direct quotes from the dialogue (no summarization or interpretation).
- Be brief but contextually complete.
- Highlight key phrases, contextual information, emotional tones, or symptoms related to the topic.

Reasoning must:
- Links the clues directly to the topic.
- Explains the logical connection between the clues and topic.
- Avoids adding external context or information not present in the clues.

**User:**
Your task is to identify important topics for the given interview.
Each topic should be concise, meaningful, and specific. Avoid combining distinct ideas or using vague terms.
Step 1 Extract CLUES.
Step 2 Generate REASONING.
Step 3 Identify TOPICS: Based on the dialogue, clues, and reasoning, identify all applicable topics.

### Output Format ###
For EACH identified topic, provide the following EXACTLY:
Identify topic: [Insert topic here]
Clues (max 200 words): [Insert clues here]
Reasoning (max 150 words): [Insert reasoning here]

Dialogue: {dialogue}

2. **Prompt for initial clue instruction (*Clue-LLM*)**

**System:**
You are a qualitative research expert with extensive experience analyzing patient-doctor dialogues. These patients are diagnosed with heart failure. You are trying to understand patient's perception of the intensity of heart failure medications. Your task is to extract key clues (limit to 200 words) directly from original dialogues supporting each given identified topic.

Clues must:
- Be direct quotes from the dialogue (no summarization/interpretation/explanation).
- Be brief but contextually complete.
- Highlight key phrases, contextual information, emotional tones, or symptoms related to the topic.

**User:**
List clues (i.e. key phrases, contextual information, semantic and emotional tones, temporal information, symptom descriptions) in the following patient-doctor dialogue that support each given identified topic.
Dialogue: {dialogue}
Topics: {topics}
### Output Format EXACTLY following ###
Topic: clues

3. **Prompt for initial reasoning instruction (*Reasoning-LLM*)**

**System:**
You are a qualitative research expert tasked with analyzing patient-doctor dialogues. These patients are diagnosed with heart failure. You are trying to understand patient's perception of the intensity of heart failure medications. Your goal is to provide a clear and concise reasoning process (limit to 150 words) based on provided clues to explain each corresponding identified topic.

Ensure your reasoning:
- Links the clues directly to the topic.
- Explains the logical connection between the clues and topic.
- Avoids adding external context or information not present in the clues.

**User:**
Based on the given clues, generate the reasoning process that supports the identified topics.
Clues: {clues}
Topics: {topics}
### Output Format EXACTLY following ###
Topic: reasoning

## 4. Prompt for evaluation (*Evaluation-LLM*)

**System:**
You are an evaluation expert tasked with analyzing a BATCH of patient-doctor dialogue clue-reasoning-topic pairs. These patients are diagnosed with heart failure, and the focus is on understanding their perception of the intensity of heart failure medications.
Your tasks are as follows:
1. Evaluate each Clues-Reasoning-Topic pair for Clue Quality and Reasoning Quality.
2. Provide feedback on both the relevance and completeness of the clues, and the logical coherence of the reasoning.
3. Identify common issues across all pairs in clue and reasoning generation.
4. Suggest improvements to the clue and reasoning prompts based on recurring patterns of errors.

**User:**
Clues-Reasoning-Topic Pair {i + 1}
**Clues:** {clues}
**Reasoning:** {reasoning}
**Topics:** {topics}

### Evaluation Task ###
For the above Clues-Reasoning-Topic pair:
1. **Clue Quality:** Evaluate the clues based on the following:
- How relevant and accurate are the clues in supporting the topic(s)?
- Are the clues complete (include all key information) and free of irrelevant details?
2. **Reasoning Quality:** Assess the reasoning based on the following:
- Does the reasoning logically connect the clues to the topic(s)?
- Are there any gaps or missing logic in the reasoning process?
- Is the reasoning concise and free of unnecessary content?

### Aggregate Feedback Task ###
Based on your evaluation of all the Clues-Reasoning-Topic pairs, provide in the following format EXACTLY following:
### Aggregate Feedback Task ###
**Common Issues:**
- **Clue Generation:** Identify recurring problems in the generated clues (e.g., missing context, irrelevant clues).
- **Reasoning Generation:** Highlight frequent issues in reasoning (e.g., logical gaps, weak connections between clues and topics).
**Suggestions for Improvement:**
- **Clue Prompt:** Propose specific improvements to the clue generation prompt.
- **Reasoning Prompt:** Recommend actionable enhancements to the reasoning generation prompt.

## 5. Prompt for optimization (*Optimization-LLM*)

**System:**
You are part of an optimization system that improves text. You will be asked to creatively and critically improve the clue prompt and reasoning prompt (instructions). You will receive some feedback, and use the feedback to improve both clue and reasoning prompts simultaneously. The feedback may be noisy, identify what is important and what is correct. Pay attention to the role description of the clue and reasoning prompts (instructions), and the context in which it is used.

**User:**
You are tasked with improving two prompts based on provided feedback.

### Task Description ###
Given the following feedback, improve both the **Clue Prompt** and the **Reasoning Prompt** simultaneously to address the issues and suggestions provided:
1. The **Clue Prompt** should:
- Guide the user/system to extract relevant, precise, and contextually complete clues directly from the dialogue.
- Ensure the clues are accurate quotes, avoid irrelevant or incomplete clues, and incorporate missing elements identified in the feedback.
- Focus on ensuring clarity and usability of the prompt.
2. The **Reasoning Prompt** should:
- Guide the user/system to logically and effectively connect the clues to the identified topics.\n"
- Ensure the reasoning structure is clear, addresses logical gaps, and builds a strong link between the clues and topics.\n"
- Incorporate improvements to reasoning clarity and structure as per the feedback.

### DO NOT VIOLATE THE FOLLOWING SYSTEM RULES ###
These are foundational instructions that must NEVER be contradicted or weakened.
[CLUE RULES]
Be direct quotes from the dialogue (no summarization/interpretation/explanation).
Quotes must be brief but contextually complete.
[REASONING RULES]
Use only the provided clues.
Do not introduce external information or assumptions.

### Provided Inputs ###
**Feedback:**{feedback}
**Current Clue Prompt:**{clue_prompt}
**Current Reasoning Prompt:**{reasoning_prompt}

### Output Instructions ###
You MUST provide your improved prompts formatted as follows:
- For the clue prompt: <IMPROVED_CLUE_PROMPT> your improved clue prompt text </IMPROVED_CLUE_PROMPT>
- For the reasoning prompt: <IMPROVED_REASONING_PROMPT> your improved

reasoning prompt text </IMPROVED_REASONING_PROMPT>
The text provided between these tags will directly replace the current prompts, so ensure your improvements are complete, clear, and directly address the feedback provided.

6. **Prompt for code identification (*Thematizer*)**

**System:**
### Background ###
Heart failure patients' perception of medication intensity is a complex experience influenced by factors beyond dosage, including side effects, treatment burden, psychological impact, quality of life, and cost. This perception is shaped by patient characteristics such as age, disease severity, comorbidities, gender, and socioeconomic status. Understanding these perceptions, along with patients' beliefs about their medications, is crucial for improving adherence and tailoring treatment strategies to enhance their quality of life. Ultimately, a patient-centered approach that involves effective communication and shared decision-making is essential in managing heart failure medication regimens.

You are a qualitative research expert tasked with identifying topics of patient-doctor dialogues. These patients are diagnosed with heart failure. You are trying to understand patient's perception of the intensity of heart failure medications.

Your task:
- Identify **all applicable topics** for the given dialogue (there may be more than one).
- For each identified topic, provide clues and reasoning to explain the connection.

Clues must:
- Be direct quotes from the dialogue (no summarization or interpretation).
- Be brief but contextually complete.
- Highlight key phrases, contextual information, emotional tones, or symptoms related to the topic.

Reasoning must:
- Links the clues directly to the topic.
- Explains the logical connection between the clues and topic.
- Avoids adding external context or information not present in the clues.

**User:**
Your task is to identify ALL applicable topics for the given dialogue. Topics should be about patient's perception of the intensity of heart failure medications. There may be multiple topics, so ensure you capture each distinct one.

Step 1 Extract CLUES: {optimized_clue_prompt}
Step 2 Generate REASONING: {optimized_reasoning_prompt}

Step 3 Identify TOPICS: Based on the dialogue, clues, and reasoning, identify all applicable topics.

### Output Format ###
For EACH identified topic, provide the following EXACTLY:
Identify topic: [Insert topic here]
Clues (max 200 words): [Insert clues here]
Reasoning (max 150 words): [Insert reasoning here]
Dialogue: {dialogue}

**7. Prompt for aggregation**

**System:**
You are an AI assistant analyzing topic identification outputs to aggregate results effectively.

**User:**
You are analyzing topic identification outputs from multiple analyses of the same patient-doctor dialogue.
### Your Goal ###
Aggregate all topics found across multiple outputs. If a topic appears in multiple outputs:
- Merge all its associated clues (without modification).
- Concisely summarize the reasoning.
- Choose the best topic name from the outputs.
For unique topics (appearing only once):
- Keep them as they are.

### Output Format ###
Provide your results in the following format for EACH topic:
Topic: [Insert best topic name]
Clues (max 200 words): [Insert aggregated clues]
Reasoning (max 150 words): [Insert summarized reasoning]
Output 1: {result_1}
Output 2: {result_2}
Output 3: {result_3}

**8. Prompt for theme generation (*CodebookGenerator*)**

**System:**
You are a qualitative research expert assisting in developing a thematic codebook from structured interview results.
Each entry consists of:
- **Topic**: an initial code identified in one interview.
- **Clues**: direct quotes from the dialogue.

- **Reasoning**: why this topic is relevant or meaningful.

### Your task ###
- Review all topic-clue-reasoning triples.
- Merge them into distinct, high-quality themes.
Each theme in the codebook should include:
- `theme_name`: The name of the higher-level concept
- `description`: A short explanation of what this theme captures and why the grouped topics fit
- `original_topics`: The list of topics it covers
- `representative_clues`: A few relevant supporting quotes
Only merge topics when there is a strong conceptual overlap. Be precise and avoid redundancy.

### Instructions for Codebook Formation ###
1. **Group original topics into broader, higher-level themes** based on a **single, clearly identifiable shared key concept**.
2. **Each original topic must belong to exactly one higher-level theme**. Avoid overlap or duplication.
3. **Do not merge topics solely based on vague thematic similarity**. Merging must be grounded in a specific, shared concept.
4. **Higher-level themes must be mutually exclusive**, covering distinct conceptual territories.
5. If an original topic does not share a strong conceptual basis with any others, treat it as its own higher-level theme.

**User:**
Below is a JSON file containing multiple identified topics with extracted clues and generated reasoning.
**Your task is to create a robust and conceptually sound codebook.** This is a crucial step in thematic analysis for organizing and synthesizing qualitative data.

### Input JSON: {original json file}

**B. Survey to understand junior investigators' perception about CoTI's outputs**

| **Output** | **Dimension** | **Question** | **Scale** |
|---|---|---|---|
| Clues | Accuracy | Is this quote directly from the transcript and correctly cited? | Yes/No |
| Clues | Relevance | Is this quote clearly related to the identified code? | 1-5 |
| Reasoning | Clarity | Is the explanation easy to follow? | 1-5 |
| Reasoning | Reasoning | Does it logically connect the clues to the code? | 1-5 |
| Code | Accuracy | Is the identified code correct based on the context? | 1-5 |
| Code | Relevance | Is the identified code related about patient perceptions of the intensity of HF medications? | 1-5 |
| Code | Comprehensiveness | Did the model miss other major codes? | Yes/No (with comment) |
| Overall | Trust | Would you trust this app to assist in annotation? | Yes/No<br>(with comment) |
| Overall | Satisfaction | Are you satisfied with this app? | 1-5 |

## C. Themes Analysis Results

### 1. Initial code discovery

| Interview | Initial Codes | Clues | Reasoning |
|---|---|---|---|
| 11 | Perceived Efficacy of Entresto | "I guess it is."<br>"No, I don't [feel much different]."<br>"If the doctor stopped the medication, is it okay?<br>"Well, I guess, they would know best, I guess. I don't know." | The interviewee repeatedly expresses uncertainty about Entresto's effectiveness, stating no noticeable symptom improvement and deferring to the doctor's judgment. This highlights her perceived lack of efficacy, a key aspect of medication intensity perception. |
| 11 | Dosage Intensity and Side Effects | "From forty-nine to ninety-seven [mg]. That was a big jump."<br>"The blood pressure increased more."<br>"Entresto makes her drowsy."<br>"They had her on a low dose at | The dosage increase caused adverse outcomes (e.g., rising blood pressure) and side effects like drowsiness, |
| | | first… then it went from what, forty something to ninety something. There was a big jump." | directly linking dosage intensity to negative physical experiences and patient concern. |
| 11 | Medication Burden from Polypharmacy | "She takes too much medication. It gets confusing."<br>"Seven at one time in the morning, three or four in the afternoon, and five at night."<br>"You've got to take the vitamins… aspirin… iron tablets."<br>"I take too many." | The daughter and patient explicitly state that the high number of medications (polypharmacy) causes confusion, cognitive burden, and overwhelm, a core theme of perceived medication intensity. |
| 11 | Cost Concerns of Entresto | "The deductible… $7,500.00… it's crazy."<br>"The only thing, back when- because he had me on a low dose… the cost?"<br>"Entresto… the only thing [financial]… a big jump." | The patient and daughter emphasize the financial strain of Entresto's high deductible and dosage changes, reflecting cost as a significant component of perceived medication intensity. |
| 11 | Impact of Side Effects on Daily Function | "The Entresto makes her drowsy."<br>"Three things [pain meds] that was taking her to sleepy, drowsy mode."<br>"I don't have any energy to do anything… half of the day is gone."<br>"It blurs my vision sometimes." | Side effects like excessive drowsiness and vision issues directly disrupt the patient's daily functioning, illustrating how perceived medication intensity affects quality of life. |
| 11 | Adherence Challenges Due to Complexity and Fatigue | "Sometimes I forget."<br>"It depends on my situation… sometimes I do, but then if it says eat first, then sometimes I don't want to be eating."<br>"She may skip the afternoon… take the night time before she goes to bed."<br>"The only thing hard for her is to prepare the meals… she has to have the walker." | The patient's forgetfulness, reliance on caregivers for organization, and physical/mental fatigue (from cancer treatment and mobility issues) create barriers to adherence, underscoring medication regimen complexity as a key perception factor. |
| 12 | High medication burden due to quantity and frequency | "I think I'm over taking them, too much medicine, but that's just me."<br>"I don't think I take that or not."<br>"About eight or nine in the morning… probably about six at night." | The patient explicitly states feeling overwhelmed by the number of medications, noting the high quantity (8–9 pills in the |
| | | "The only problem I have is being on dialysis and stuff, my bedroom is stack full of the boxes."<br>"I just … I put everything in the Lord's hands. The Lord's taking care of everything." | morning, 6 at night) and the physical clutter of medical supplies. His acknowledgment of feeling "overmedicated" and reliance on religious coping to manage the burden directly links to his perception of medication intensity as physically and emotionally taxing. |
| 12 | Financial burden of medication costs and hospitalizations | "Brilinta's, I think that's $300-and some for three-month supply."<br>"I probably spend close to $300 a month on medication."<br>"every time I go to the emergency room, I think it's … 250 out of me right there. And then if I take the ambulance, that's another 250."<br>"it ain't long and then you got $3,000, $4,000, $5,000 worth of debt."<br>"you work all your life and then to have something like this or something like that … it's depressing." | The patient emphasizes the financial strain of medications (e.g., Brilinta) and hospitalizations, including ER and ambulance fees. He highlights how these costs contribute to stress and financial insecurity, directly tying economic factors to his perception of medication intensity as a source of burden. |
| 12 | Concerns over overmedication and side effects | "I think I'm over medicated that way"<br>"I've been bleeding internally … I think that's caused from stress."<br>"they put me on … Eliquis … wasn't very long after that I was right back in here with throwing up blood or pooping blood."<br>"if you could not remember which medication is for heart? Would that be a problem ever?" | The patient attributes internal bleeding to overmedication with blood thinners (Eliquis), expressing fear of adverse effects. His uncertainty about medication purposes and the physical consequences of treatment highlight his perception of intensity as risky and potentially harmful, amplifying his psychological and physical burden. |
| 12 | Psychological coping through religious faith and acceptance | "I put everything in the Lord's hands. The Lord's taking care of everything."<br>"Isaiah 53:5 says by his stripes we are healed and I just, I'm healed. And I believe in that healing and I'm healed."<br>"if I go home to be with the Lord, it's | The patient repeatedly cites religious beliefs as a mechanism to cope with his medical challenges, framing his acceptance of medications and mortality as part of his |
| | | a winning situation also." "sooner or later everybody's going to be there anyway." | spiritual faith. This reflects how his psychological resilience shapes his perception of medication intensity as manageable despite the burden. |

## 2. Codebook generated by CoTI

| **Theme name** | **Description** | **Original codes** | **Representative clues** |
|---|---|---|---|
| Medication Burden | This code captures the emotional and psychological weight associated with managing multiple medications. It reflects patients' feelings of being overwhelmed by the quantity and complexity of their regimens, which can impact adherence and overall quality of life. | "Medication Burden", "Perception of Medication Burden", "Medication Burden and Complexity", "Medication Intensity and Side Effects" | "It's just a burden to take that many pills.", "I wish I didn't take so much.", "I'm so tired of taking all this medication.", "It's hard for one person to do it." |
| Financial Burden of Medications | This code encompasses the financial challenges patients face related to their medication costs. It reflects the emotional distress stemming from high prices, insurance issues, and the impact of financial constraints on medication adherence. | "Financial Burden of Medications", "Financial Concerns Related to Medication", "Cost of Medications", "Financial Considerations and Access to Medication" | "I probably spend close to $300 a month on medication.", "You've got to make a choice whether you pay your bills or eat.", "It's been either zero or no copay." |
| Medication Adherence and Management Challenges | This code highlights the difficulties patients encounter in adhering to their medication regimens, including forgetfulness, confusion over dosing, and the need for structured support systems. It emphasizes the importance of strategies to enhance adherence and manage complex regimens. | "Medication Adherence", "Medication Adherence Challenges", "Management of Medication Regimen", "Adherence Challenges" | "I try not to forget. Sometimes I do.", "I just take them. I just, I asked the doctor what's this for.", "We started doing a pill box a couple days before we came to the hospital this time which has helped." |
| Psychological and | This code captures the | "Psychological Impact", | "It was killing my |
| Emotional Impact of Medications | emotional responses related to medication, including anxiety about adherence, feelings of isolation, and the broader psychological toll that medications can exert on patients' lives. It also encompasses the spiritual beliefs that provide coping mechanisms. | "Medication Adherence Anxiety", "Emotional Impact of Medication", "Social Isolation Due to Medication Effects" | soul.", "I feel so bad. I think about it so much.", "I couldn't even socialize with my friends." |
| Side Effects and Their Perception | This code relates to the experiences and perceptions of side effects stemming from medications. It highlights how these effects influence patients' willingness to adhere to treatment, their quality of life, and their overall perceptions of medication effectiveness. | "Medication Intensity and Side Effects" "Side Effects Perception" "Concerns about Side Effects of Medications", "Perceived Side Effects of Medications" | "They make me so sleepy.", "I think the Lisinopril would be. The coughing would be considered a side effect.", "I get a little tired of going to the bathroom sometimes." |
| Patient-Doctor Relationship and Communication | This code encompasses patients' perceptions of their interactions with healthcare providers, including concerns about communication gaps and the understanding of treatment plans. It highlights the need for effective patient education and the influence of this relationship on medication adherence. | "Concerns Over Patient Education and Understanding", "Acceptance of Medication Changes with Aging", "Patient Awareness of Heart Failure Diagnosis" | "I don't think I've ever been officially told that.", "Nobody told me.", "I need to do that." |
| Desire for Simplified Medication Regimen | This code reflects patients' preferences for a more streamlined and manageable medication regimen. It captures the sentiment that simpler regimens could enhance adherence and reduce perceived burdens associated with complex dosing schedules. | "Desire for Simplified Medication Regimen", "Desire for Reduced Medication Regimen" | "If she could take less, it would be wonderful.", "I wish I could take them all at once." |
| Perceived | This code indicates | "Perceived Medication | "As long as they keep |
| Effectiveness of Medications | patients' perceptions regarding the effectiveness of their medications and treatment plans. It highlights the emotional complexity tied to treatment efficacy, including feelings of disappointment and hope." | Effectiveness", "Perception of Treatment Effectiveness", "Concerns about Medication Efficacy" | her blood pressure normal.", "I think it's fine the way it is." |

**Table S1.** Assigned Transcripts for Junior Investigators

| Junior Investigator | Assigned Transcripts |
|---|---|
| A | 1, 5, 6, 7, 8, 9, 10 |
| B | 1, 2, 3, 4, 5, 9, 10 |
| C | 1, 2, 3, 7, 8, 9, 10 |
| D | 4, 5, 6, 7, 8, 9, 10 |

**Table S2.** Computation evaluations among three external datasets.

| | **COVID-19 Dataset** | **Strategy Dataset** |
|---|---|---|
| Clue Similarity (Recall) | 0.732 | 0.675 |
| Theme Similarity (Cosine similarity) | 0.606 | 0.511 |

**Table S3.** Comparison of themes developed by experts and CoTI (without human) on the Sierra Leone Covid-19 interview transcripts.

| **Experts** | **Overlap** | **CoTI** |
|---|---|---|
| Patient experiences<br>Information | Health workforce<br>Service delivery<br>Availability of medicines & supplies<br>Leadership and governance<br>Financing<br>Learning from Ebola | Public Health Communication and Awareness<br>Patient Access and Public Perception |

**Table S4.** Comparison of themes developed by experts and CoTI (without human) on the family carers' strategies interview transcripts.

| **Experts** | **Overlap** | **CoTI** |
|---|---|---|
| Keeping a routine<br>Providing comfort and reassurance<br>Activities and interests<br>Being calm, respectful and avoiding arguments<br>Allowing the behaviors to continue<br>Walking away<br>Time for themselves | Working out and responding to the cause of agitated behaviors<br>Communication<br>Keeping people safe<br>Emotional support<br>Paid support<br>Practical support<br>Understanding agitation as part of dementia | Caregiver strategies for agitation management<br>Support systems and resources for caregivers<br>Behavioral changes in dementia patients<br>Anticipatory grief and loss |